\documentclass[pdftex]{vienna-conf2019}              
\usepackage{graphicx}
\usepackage{hyperref}
\usepackage[]{natbib}  
\usepackage{epstopdf}

\usepackage{upgreek}
\usepackage{amsmath,amssymb}
\usepackage{booktabs,caption,fixltx2e}	

\def\BibTeX{{\rm B\kern-.05em{\sc i\kern-.025em b}\kern-.08em
    T\kern-.1667em\lower.7ex\hbox{E}\kern-.125emX}}
\bibpunct{(}{)}{;}{a}{}{,}  

\begin{document}

\TitreGlobal{Stars and their variability observed from space}


\title{Search for quiet stellar-mass black holes\\
by asteroseismology from space}

\runningtitle{Search for quiet stellar-mass black holes by asteroseismology from space}

\author{H. Shibahashi}\address{Department of Astronomy, School of Science, University of Tokyo, Tokyo 113-0033, Japan}

\author{S. J. Murphy}\address{Sydney Institute for Astronomy, School of Physics, University of Sydney, NSW 2006, Australia}




\setcounter{page}{1}


\maketitle


\begin{abstract}
Stars with an initial mass more than $\sim 25\,{\rm M}_\odot$ are thought to ultimately become black holes. Then stellar-mass black holes should be ubiquitous but fewer than 20 have been found in our Galaxy to date, all of which have been found through their X-ray emission. In most cases these are soft X-ray transients --- low-mass X-ray binaries whose optical counterparts are late type stars filling their Roche lobes, leading to accretion onto black holes. In one case, the stellar-mass black hole is in a high-mass X-ray binary whose optical counterpart is an early type star. Its strong stellar winds are accreted by the black hole, producing X-ray emission.
It follows that X-ray-quiet stellar-mass black holes exist in wide binary systems. The discovery of black holes in the optical through their gravitational interactions would be a major scientific breakthrough. Recent space-based photometry has made it possible to measure phase or frequency modulation of pulsating stars to extremely high precision. Such modulation is caused by orbital motion, and its analysis offers the lower limit for the mass of the companion to the pulsating star. If the companions are non-luminous and if the masses of the companions exceed the mass limit for neutron stars ($\sim 3 \,{\rm M}_\odot$), the companions should be black holes. We review the methodology, and analyses of some encouraging cases are demonstrated.
\end{abstract}



\section{Introduction}
It is thought that stars with an initial mass more than $\sim 25\,{\rm M}_\odot$ ultimately end their lives as black holes. Though the mass thresholds are dependent on the metallicity of the stars and are currently uncertain, stars with an initial mass in the range of $\sim 25$--$40\,{\rm M}_\odot$ are expected to form black holes following a supernova explosion. Even more massive stars collapse to black holes without spectacular explosion \citep{Heger_et_al_2003}. These events are thought to correspond to the observed phenomena known as ``failed supernovae'', in which sudden brightening occurs as in the early stage of a supernova, but then does not develop to full supernova luminosity \citep{Adams_et_al_2017}. 
Hence stellar-mass black holes should be ubiquitous. 
Population statistics of these stellar-mass black holes in our Galaxy may be estimated with a reasonable initial mass function, a star formation rate, and using assumptions of the density structure of the Galaxy. Though the uncertainty is large, it is estimated that more than 100 million black holes reside in our Galaxy \citep{Brown_Bethe_1994, Mashian_Abraham_2017, Breivik_et_al_2017, Lamberts_et_al_2018, Yalinewich_et_al_2018, Yamaguchi_et_al_2018}. 

Since black holes emit no light, there is no way to see them directly. However, if a black hole is in a close binary system with an ordinary star, its presence can manifest itself in X-ray emission as follows: Sometimes the black hole swallows gas from the companion star. As this gas swirls around the black hole, a huge amount of potential energy is liberated, ultimately emitting X-rays that are detectable from space. 
However, X-ray binaries are not limited to systems containing a black hole. In general, they are close binaries composed of a compact object, either a neutron star or a black hole, which is accreting mass from a companion donor star. 
Spectroscopic radial velocity measurements of the optically visible donor star allow us to deduce the lower limit of the invisible compact object of the binary system, even if the latter is not visible itself.  According to current theory, a neutron star more massive than $3\,{\rm M}_\odot$ is unstable and will collapse into a black hole. Therefore, a mass function that corresponds to a companion exceeding this critical mass is considered to be reliable evidence for a black hole.
This is how stellar-mass black holes have so far been detected and confirmed (see reviews, e.g. \citealt{Cowley_1992, Remillard_McClintock_2006, Casares_Jonker_2014}, and references therein). But more importantly, despite the aforementioned estimates of $10^8$ black holes in our Galaxy, fewer than 20 stellar-mass black holes have so far been confirmed \citep{Corral-Santana_et_al_2016, Torres_et_al_2019}. 
The X-ray emissions resulting from accretion may occur only in a close binary system.
Hence, it is natural to expect quiescent stellar-mass black holes, without X-ray emission, in binary systems with large separations.

\section{Black-hole X-ray binaries}
\begin{figure}[ht!]
 \centering
 \includegraphics[width=0.6\textwidth,clip]{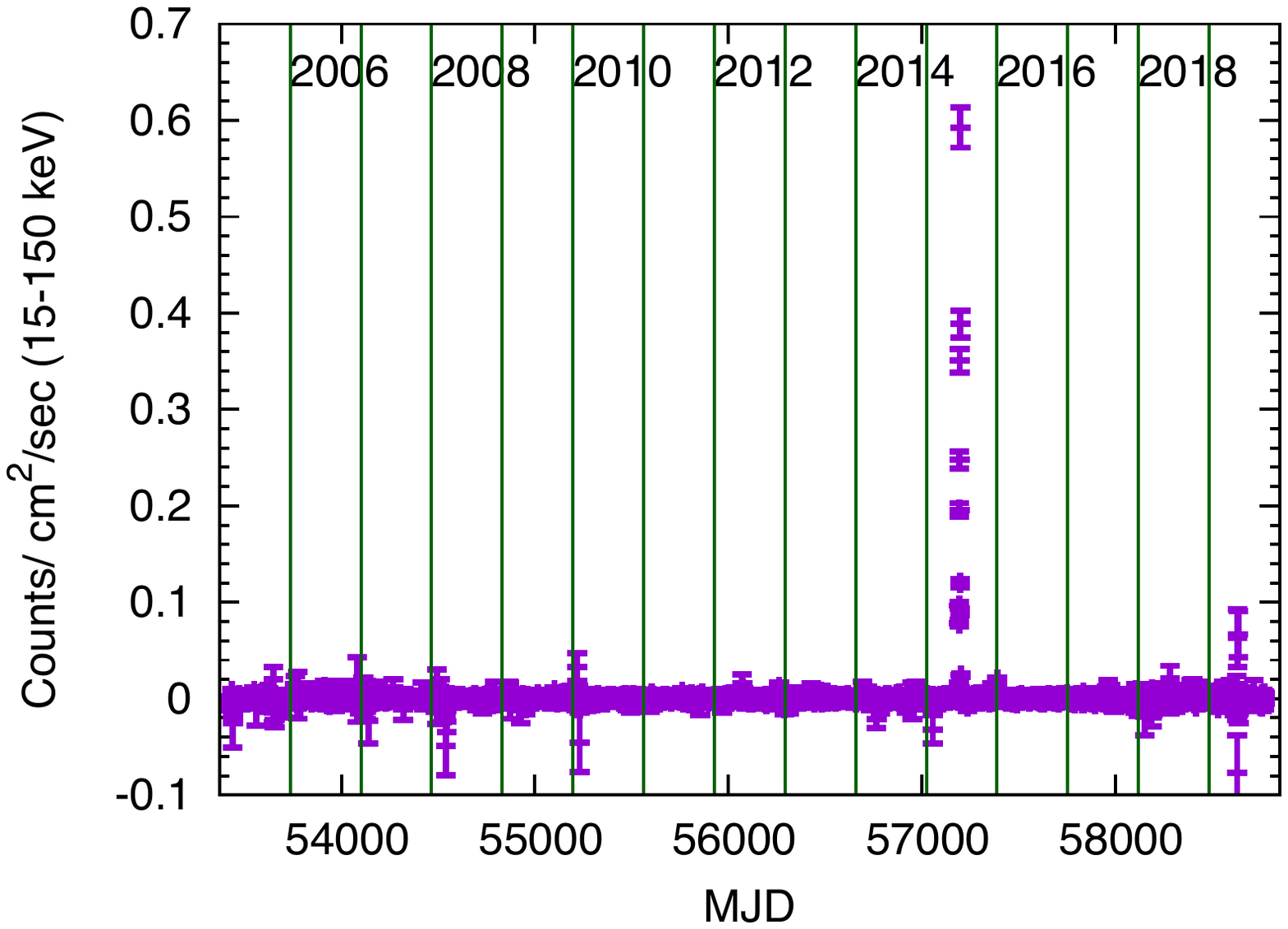}       
  \caption{A $\gamma$-ray light curve of GS\,2023+338 obtained by the Burst Alert Telescope (BAT) onboard the {\it Swift} satellite. The time series data are taken from the Swift data archives.} 
  \label{shibahashi:fig01}
\end{figure}

About one third of X-ray binaries are not persistently visible but are detected as transient sources. 
The Galactic stellar-mass black holes have so far been found, except for Cyg X-1, as transient X-ray sources in binaries with low-mass K or M dwarf counterparts \citep{Tanaka_Shibazaki_1996}. 
Figure\,\ref{shibahashi:fig01} shows the $\gamma$-ray light curve, obtained by the Burst Alert Telescope (BAT) on the Neil Gehrels Swift Observatory ({\it Swift} satellite) \citep{Swift_2004, BAT_2005}, of such a low-mass X-ray binary, a {\it Ginga} source GS\,2023+338. 
Though the light curves are different from source to source, these systems are often called X-ray novae due to their X-ray brightening. The outbursts are caused by mass transfer instabilities in an accretion disc, which is fed by a low-mass dwarf star \citep{Mineshige_Wheeler_1989}.
During the $\gamma$-ray (and X-ray) burst in 2015, the optical counterpart known as V404 Cyg brightened by $\sim 6$\,mag in $V$ band. Such optical outbursts, first notified as Nova Cygni 1938 by A. A. Wachmann, seem to be caused by reprocessing of X-ray photons in the accretion disk. In some other X-ray transients, the sudden optical brightening was also classified as a nova, like Nova Vel 1993 in the case of GRS\,1009$-$45. 
The naming may be misleading, since the physical mechanism of these systems are different from that for classical novae, which are caused by a thermonuclear flash triggered by the accumulation of accreted gas on the surface of white dwarfs.

In the X-ray quiescent phase between outbursts of GS\,2023+338, V404 Cyg is as faint as $V\sim 18\,{\rm mag}$, and the optical source is the low-mass donor star itself \citep{Casares_et_al_2019}, hence radial velocity measurements become possible and the orbital elements are deduced from them.
The mass function giving the lower limit of the invisible compact object is determined by the orbital period $P_{\rm orb}$, the amplitude of radial velocity variation $K_{\rm opt}$ and the eccentricity $e$:
\begin{eqnarray}
	f(M_{\rm opt},M_{\rm X},\sin i) &:=& {{M_{\rm X}^3\sin^3 i}\over{(M_{\rm opt}+M_{\rm X})^2}}
	\nonumber\\
	&=&
	{{1}\over{2\uppi G}}
	P_{\rm orb} K_{\rm opt}^3 \left(1-e^2\right)^{3/2},
\label{eq:2.1}
\end{eqnarray}
where $M_{\rm X}$ and $M_{\rm opt}$ denote the masses of the X-ray emitting, but optically invisible, compact object and of the optical counterpart, respectively, $i$ is the inclination angle, and $G$ is the gravitational constant.

For V404 Cyg, the orbital period is 6.5\,d, and the amplitude of (almost sinusoidal) radial velocity variation is $\sim 200\,{\rm km}\,{\rm s}^{-1}$, so the mass function is $\sim 6\,{\rm M}_\odot$, giving a companion mass $M_{\rm X}$ much larger than the critical mass for a neutron star. Hence, this binary system contains a stellar-mass black hole \citep{Casares_et_al_1992}.

\begin{figure}[t]
 \centering
 \includegraphics[width=0.6\textwidth,clip]{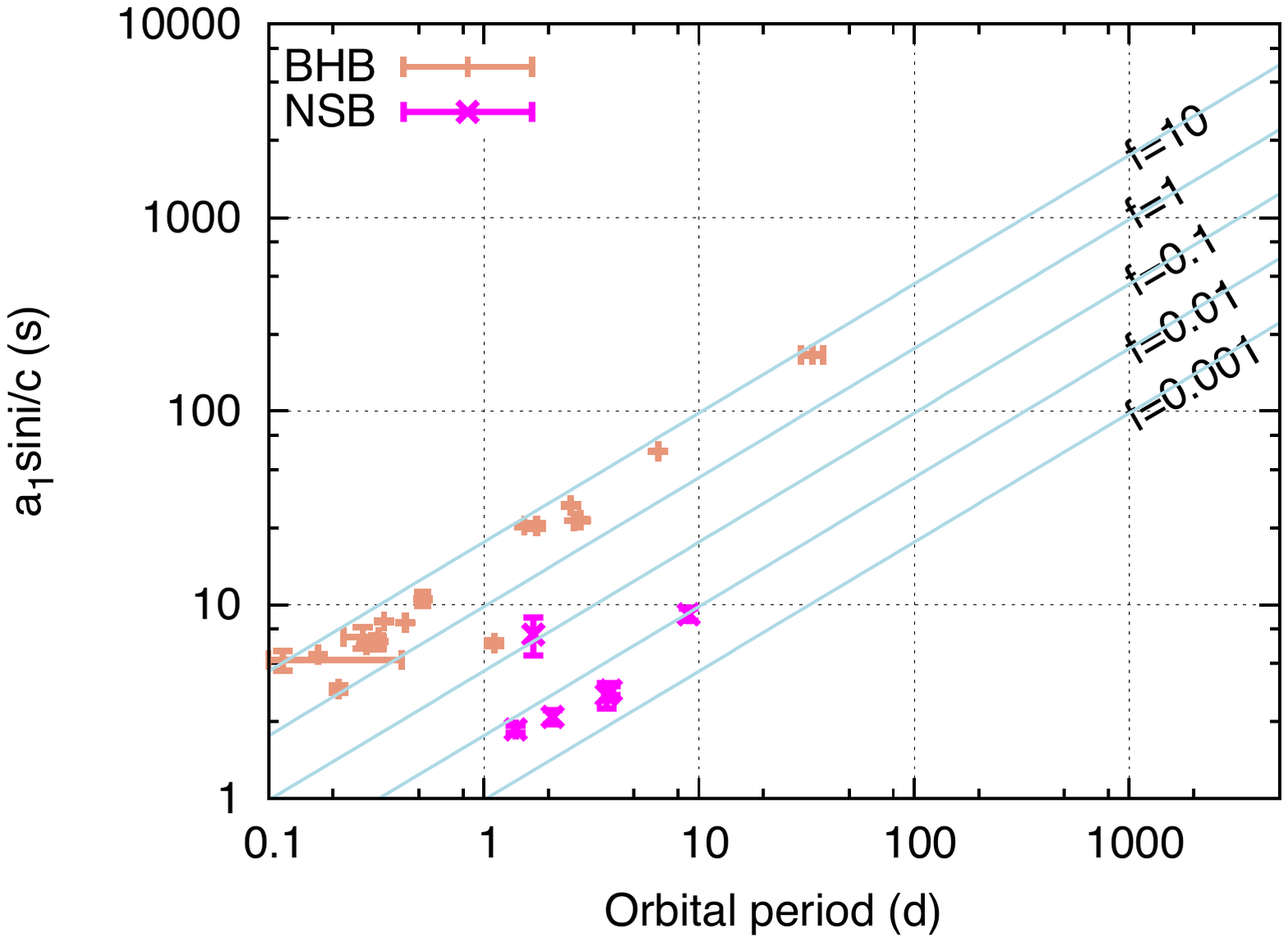}      
  \caption{Orbital period vs projected semi-major axis of X-ray binary systems.
  Mass functions are shown in units of ${\rm M}_\odot$.}
  \label{shibahashi:fig02}
\end{figure}

Besides GS\,2023+338, sixteen X-ray binaries have been confirmed, from spectroscopic radial velocities of the optical counterparts, to be composed of a late-type low mass star and a stellar-mass black hole. 
In addition,
the presence of a stellar-mass black hole has been confirmed in Cyg X-1 whose optical counterpart is an early type star \citep{Webster_Murdin_1972, Bolton_1972}. Its strong stellar winds are accreted by the black hole, producing X-ray emission.
The binary properties of all of these Galactic black holes in the literature are listed in Table\,\ref{table:01} (\citealt{Corral-Santana_et_al_2016} and references therein).
For each X-ray binary system, the projected semi-major axis of the optical counterpart, in units of light-seconds, is plotted versus the orbital period in Fig.\,\ref{shibahashi:fig02}, where both axes use a logarithmic scale. Here, the semi-major axis is 
\begin{equation}
	{{a_{\rm opt}\sin i}\over{c}} = {{1}\over{2\uppi c}}P_{\rm orb} K_{\rm opt}\left(1-e^2\right)^{1/2},
\label{eq:2.2}
\end{equation}
where $c$ is the speed of the light.
Systems of the same mass function would form a line with an inclination of 2/3. 
In the same diagram, some representative data for neutron-star X-ray binaries are plotted. As expected, most of the black-hole binaries (BHB) have a mass function larger than 1\,${\rm M}_\odot$, while the neutron-star binaries (NSB) have substantially smaller values.

\begin{table}[t]
\caption{Binary properties of the dynamically confirmed black-hole X-ray binaries in our Galaxy.}
\begin{center}
\begin{tabular}{cccccc}
\toprule
X-ray source & Opt. & Sp. Type & $P_{\rm orb}$ (h) & $f(M)$ (M$_\odot$) & $M_{\rm BH}$ (M$_\odot$) \\
\midrule
Cyg X-1 & HDE226868 & O9Iab & 134.4 & $0.244 \pm 0.006$ & $14.8 \pm 1.0$ \\
GROJ0422$+$32 & N. Per & M4-5V & 5.09 & $1.19 \pm 0.02$ & 2 -- 15 \\
3A0620$-$003 & N. Mon & K2-7V & 7.75 & $2.79 \pm0.04$ & $6.6 \pm 0.3$ \\
GRS1009$-$45 & N. Vel 93 & K7-M0V & 6.84 & $3.2 \pm 0.1$ & $ > 4.4$ \\
XTEJ1118$+$480 &  & K7-M1V & 4.08 & $6.27 \pm 0.04$ & 6.9 -- 8.2 \\
GRS1124$-$684 & N. Mus 91 &  K3-5V & 10.38 & $3.02 \pm 0.06$ & 3.8 -- 7.5 \\
GS1354$-$64 & BW Cir & G5III & 61.07 & $5.7 \pm 0.3$ & $> 7.6$ \\
4U1543$-$475 & IL Lup & A2V & 26.79 & $0.25 \pm 0.01$ & 8.4 -- 10.4 \\
XTEJ1550$-$564 & & K2-4IV & 37.01 & $7.7 \pm 0.4$ & 7.8 -- 15.6 \\
XTEJ1650$-$500 & & K4V & 7.69 & $2.7 \pm 0.6$ & $< 7.3$ \\
GROJ1655$-$40 & N. Sco 94 & F6IV & 62.92 & $2.73 \pm 0.09$ & $6.0 \pm 0.4$ \\
1HJ1659$-$487 & GX 339-4 & $>$ GIV & 42.14 & $5.8 \pm 0.5$ & $> 6$ \\
H1705$-$250 & N. Oph 77 & K3-M0V & 12.51 & $4.9 \pm 0.1$ & 4.9 -- 7.9 \\
SAXJ1819.3$-$2525 & V4641 Sgr & B9III & 67.62 & $2.7 \pm 0.1$ & $6.4 \pm 0.6$ \\
XTEJ1859$+$226 & V406 Vul & K5V & 6.58 & $4.5 \pm 0.6$ & $> 5.42$ \\
GRS1915$+$105 & & K1-5III & 812 & $7.0 \pm 0.2$ & $12 \pm 2$ \\
GS2000$+$251 & QZ Vul & K3-7V & 8.26 & $5.0 \pm 0.1$ & 5.5 -- 8.8 \\
GS2023$+$338 & V404 Cyg & K3III & 155.31 & $6.08 \pm 0.06$ & $9.0^{+0.2}_{-0.6}$ \\
\bottomrule
\end{tabular}
\end{center}
\label{table:01}
\end{table}

\begin{figure}[b!]
 \centering
 \includegraphics[width=0.6\textwidth,clip]{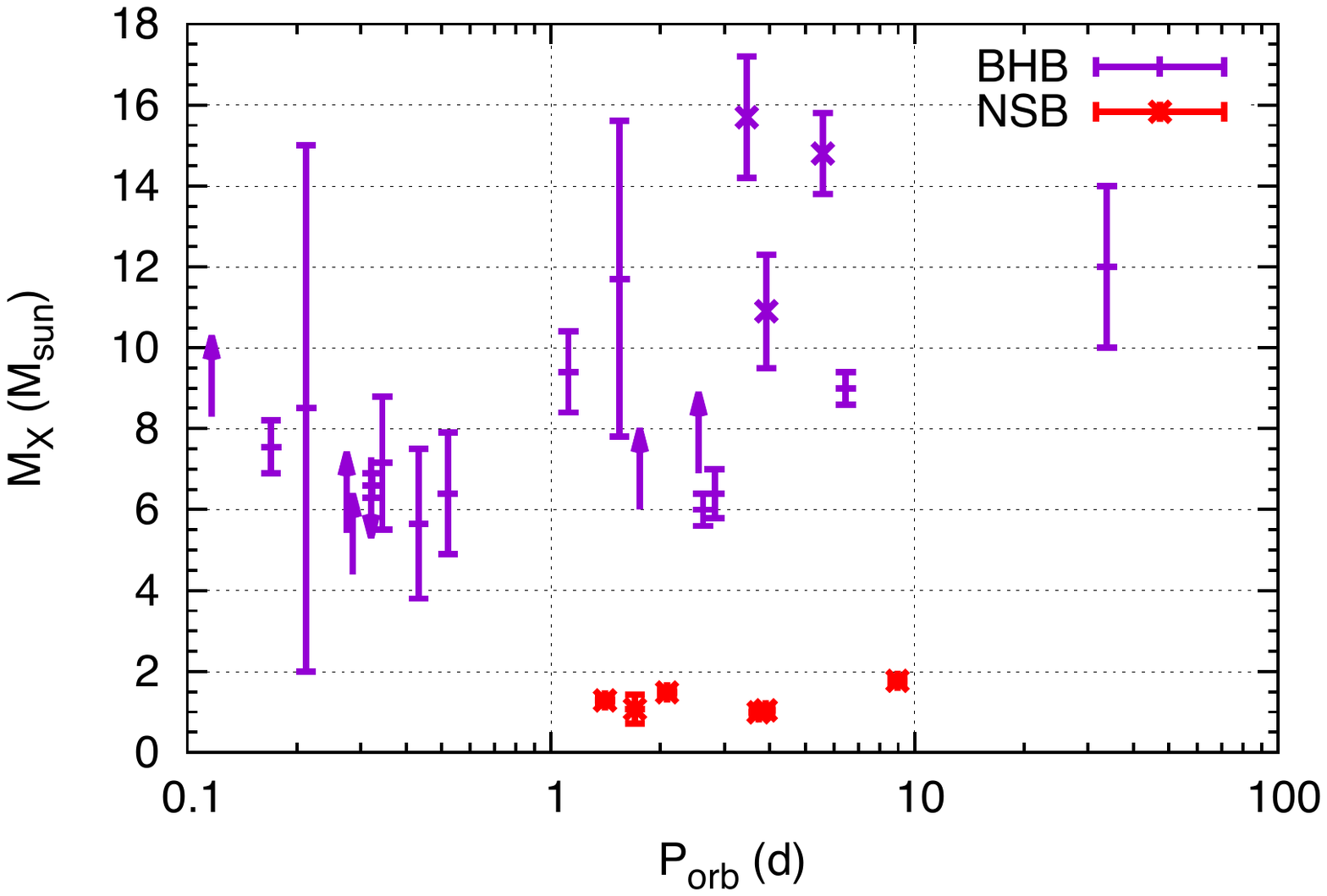}      
  \caption{Mass of the invisible compact object vs orbital period of X-ray binary systems.}
  \label{shibahashi:fig03}
\end{figure}

Since the low-mass star is tidally locked, the spin period is the same as the orbital period. Hence the rotation velocity and the orbital velocity are different only by the factor of the ratio of the size of the star and the size of the orbit. The star fills its Roche lobe, so the size of the star is the Roche lobe size. The Roche lobe size relative to the binary separation is determined by the mass ratio \citep{Paczynski_1971}. Hence, the ratio of the rotation velocity to the orbital velocity gives the mass ratio. With a reasonable estimate for the mass of the low-mass star from its spectrum, the mass of the invisible object is thus observationally determined.
The results taken from the literature are illustrated in Fig.\,\ref{shibahashi:fig03} (\citealt{Corral-Santana_et_al_2016} and references therein). The X-ray objects with an estimated mass in the range of $1.5$--$3\,{\rm M}_\odot$ are neutron stars. The X-ray transient systems with an invisible object with a mass higher than $3\,{\rm M}_\odot$ are considered as the best signature of stellar-mass black holes.

X-ray binaries containing black holes are mostly distinguishable from those with neutron stars by their X-ray spectrum. The former are characterised by an ultrasoft thermal component ($\lesssim 1.2\,{\rm keV}$), accompanied by a hard tail and seen after the flux reaches to the maximum, while the latter show a little bit harder ($\sim$ a few keV) blackbody component, which is thought to be from the neutron star envelope, and a softer but still a bit harder component than the black-hole X-ray binaries, most probably from the accretion disk. 
Ultrasoft X-ray transient sources are regarded as a signature of binaries containing a black hole.
There are $\sim 60$ such X-ray sources suspected to be black hole binaries from their spectra, classified as ``black-hole candidates'', but not yet dynamically confirmed and then less secure (\citealt{Corral-Santana_et_al_2016} and references therein).

\section{Search for quiet black holes in single-lined spectroscopic binaries}
Quiescent black holes in wide binary systems could be found in radial velocity surveys by searching for single-lined systems with high mass functions. Such an attempt was first proposed by \citet{Guseinov_Zeldovich_1966}, who selected seven plausible candidates in an attempt to detect ``collapsed stars'', as they were then called, before \citet{Wheeler_1968} popularized the terminology ``black hole''. Their attempt was later followed by \citet{Trimble_Thorne_1969} (see also \citealt{Trimble_Thorne_2018}) by using the sixth catalogue of the orbital elements of spectroscopic binary systems \citep{SB6_1967}. 
For each system, the mass of the primary star was estimated from its spectral
type, and an approximate lower limit to the mass of the unseen companion was
then calculated from the observed mass function. 
\citet{Trimble_Thorne_1969} listed 50 systems having an unseen secondary star more massive than the Chandrasekhar mass limit for a white dwarf. 
\begin{figure}[ht!]
 \centering
 \includegraphics[width=0.6\textwidth,clip]{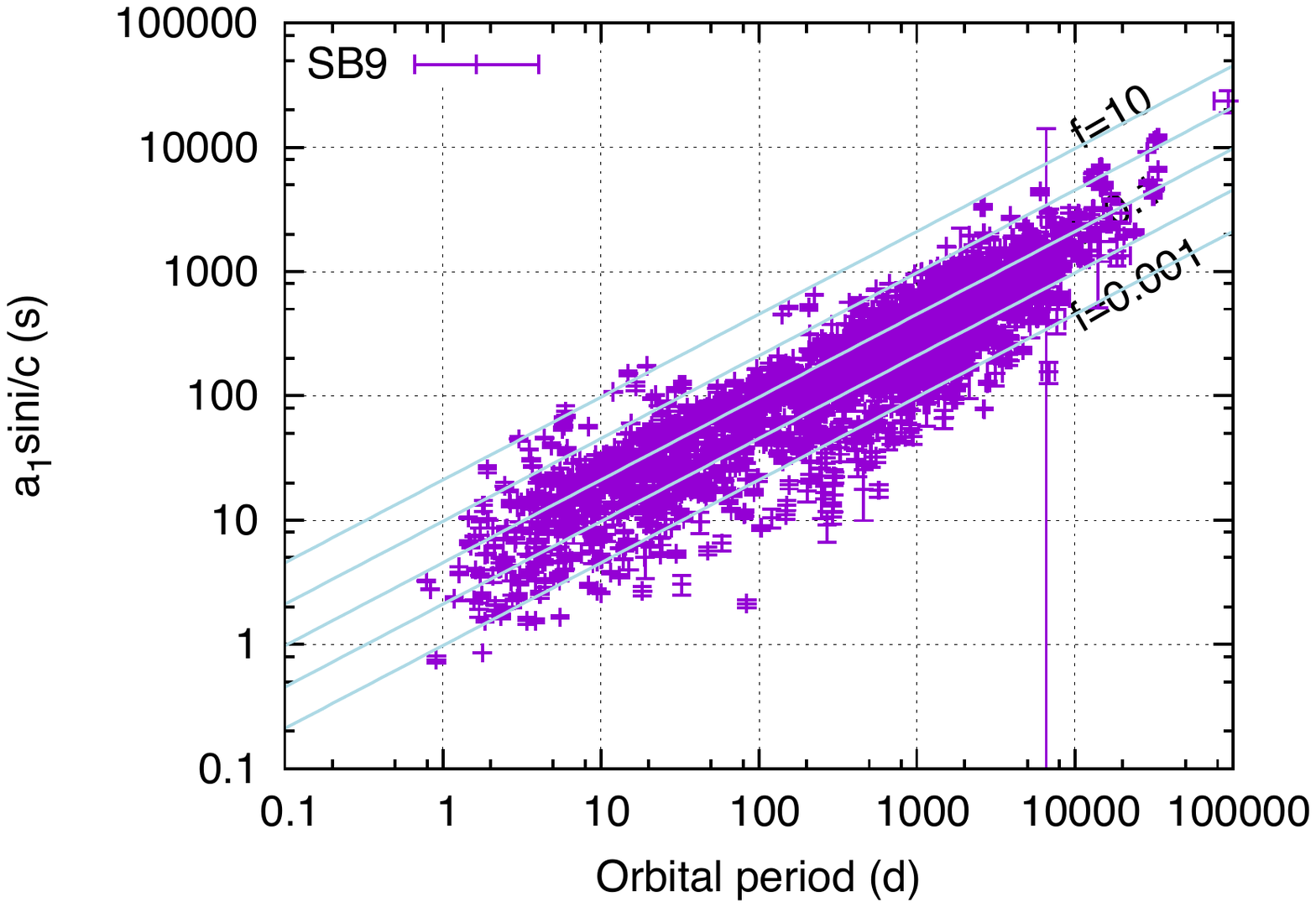}       
  \caption{Orbital period vs projected semi-major axis of binary systems listed in the ninth catalogue of spectroscopic binary systems (SB9).
  Mass functions are shown in units of ${\rm M}_\odot$.}
  \label{shibahashi:fig04}
\end{figure}

Figure\,\ref{shibahashi:fig04} shows the distribution of spectroscopic binaries listed in the latest ninth catalogue of spectroscopic binary systems \citep{SB9_2004} in the orbital period versus the semi-major axis plane.  More than fifty systems are found to have a mass function larger than 1\,${\rm M}_\odot$.
However, in most cases, careful follow-up observations unveiled the presence of a fainter secondary star of larger mass (e.g. \citealt{Stickland_1997}). 
A plausible scenario is that the primary was originally more massive than the secondary and it has evolved faster than the secondary. A substantial amount of mass loss in the red giant phase made the primary significantly less massive than the secondary, which is still in the main-sequence phase and relatively much fainter than the evolved primary. 
Generally, a larger and more homogeneous sample of main-sequence stars is favourable, however, such a sample has been difficult to procure in practice.
Spectroscopic data have to be collected as time series covering the orbital phase of each binary system, often obtained one by one from ground-based observing sites with a large amount of observing time, often on large telescopes for fainter stars. This is the bottle-neck to binary studies. 

Recently, two months after this conference, \citet{Thompson_2019} reported their finding of a black hole-giant star binary system, 2MASS J05215658+4359220, in multi-epoch radial velocity measurements acquired by the Apache Point Observatory Galactic Evolution Experiment (APOGEE) survey \citep{APOGEE_2017}. The system was found to have a nearly circular orbit with the period  $P_{\rm orb}=83.2\pm 0.06\,{\rm d}$ and the semi-amplitude $K_{\rm opt}\simeq 44.6\pm 0.1\,{\rm km\,s}^{-1}$. The mass function is then $0.766\pm 0.006\,{\rm M}_\odot$. The photometric data show periodic variation with the same period as the radial velocity variation, implying spin-orbit synchronisation of the star.
A combination of the measured projected spin velocity and the period leads to $R_{\rm opt}\sin i \simeq 23\pm 1\,{\rm R}_\odot$, and the mass of the giant star is estimated to be $M_{\rm opt}\sin^2 i \simeq 4.4^{+2.2}_{-1.5}\,{\rm M}_\odot$ from this radius and the spectroscopically estimated $\log g$. 
An independent combination of the apparent luminosity, the spectroscopically determined effective temperature and Gaia distance measurements gives $R_{\rm opt} = 30^{+9}_{-6}\,{\rm R}_\odot$, being consistent with the value based on the spin velocity. As a consequence, from the aforementioned mass function, the minimum mass of the unseen companion is estimated to be $\sim 2.9\,{\rm M_\odot}$, marginally exceeding the critical mass for a neutron star.
This is likely to be the first success of finding an X-ray quiet black hole lurking in a binary system, though within the uncertainties the companion could be a neutron star instead.

\section{Search for quiet black holes based on phase modulation of pulsating stars in binaries}
Starting with the Canadian {\it MOST} mission \citep{MOST_2003}, through the European Space Agency (ESA) mission {\it CoRoT} \citep{CoRoT_2009}, the National Aeronautics and Space Administration (NASA) space missions {\it Kepler} \citep{Kepler_2010} and {\it TESS} \citep{TESS_2015} and the international {\it BRITE}-Constellation \citep{BRITE_2014}, space-based photometry with extremely high precision over long time spans has led to a drastic change of the situation mentioned in the previous section, and   has revolutionized our view of variability of stars. Some variability has been detected in almost all stars, and tens of thousands of pulsating stars were newly discovered, along with hundreds of eclipsing binaries \citep{Kirk_et_al_2016}. 
{\it Kepler}'s four-year simultaneous monitoring of nearly 200\,000 stars also opened a new window to a statistical study of binary stars and their orbits \citep{Murphy_et_al_2018, Murphy_2018, Liege_2019}.
This pedigree will be further taken over by the ESA mission {\it PLATO}\footnote{http://sci.esa.int/plato/59252-plato-definition-study-report-red-book/}.

\begin{figure}[t]
\centering
\includegraphics[width=0.7\linewidth]{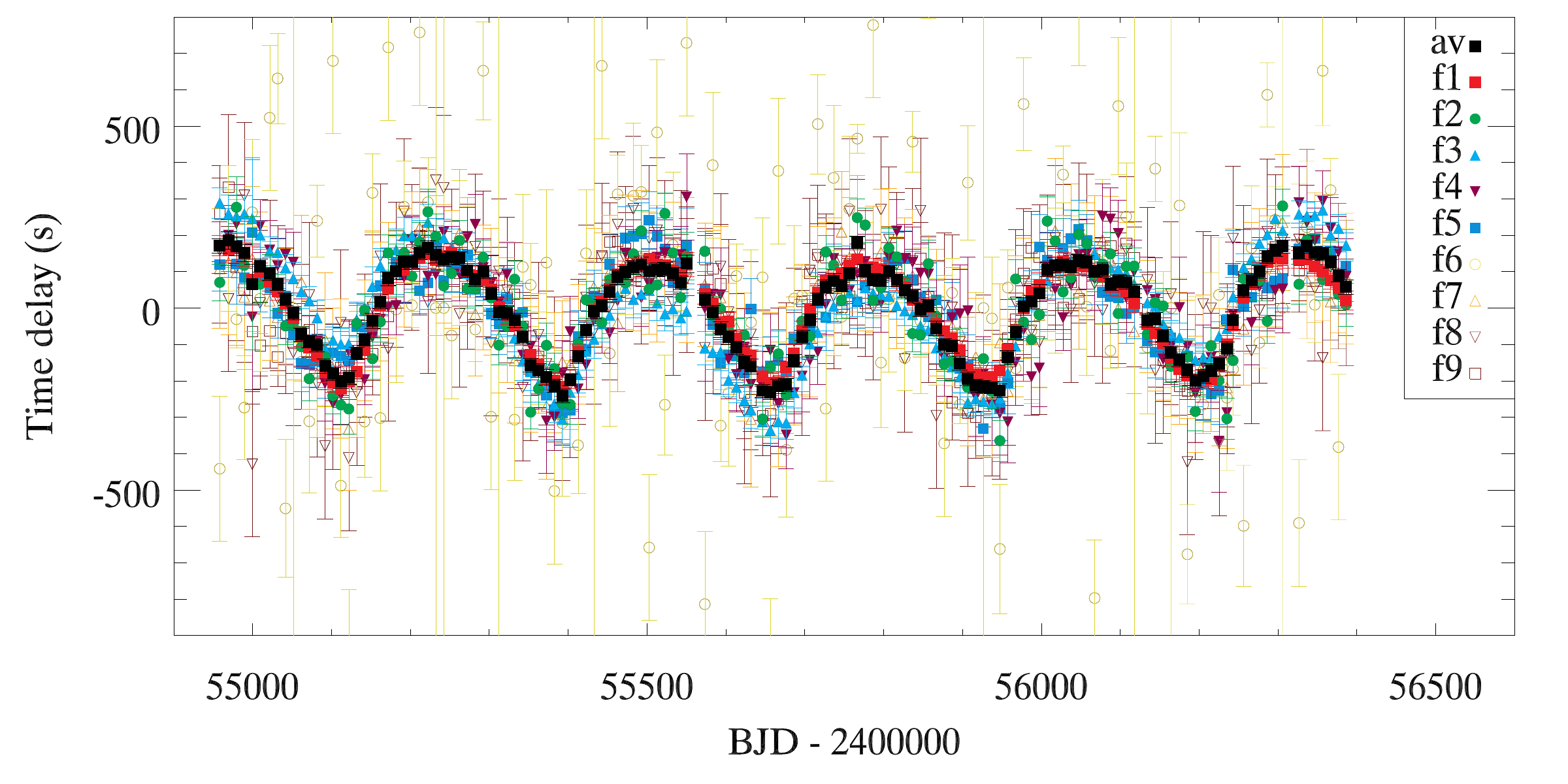} 
\caption{An example of time delay curve (KIC\,9651065) using nine different pulsation modes. The weighted average is shown as filled black squares. Adopted from \citet{Murphy_Shibahashi_2015}.
\label{shibahashi:fig05}}
\end{figure}

Binary orbital motion causes a periodic variation in the path length of light travelling to us from a star. 
Hence, if the star is pulsating, the time delay manifests itself as a periodically varying phase shift in the form of the product with the intrinsic angular frequency \citep{FM_2012, FM2_2015}. 
The light-arrival time delay is hence measurable by dividing the observed phase variation by the frequency \citep{PM_2014}. 
In the past, the light-time effect on the observed times of maxima in luminosity, which vary over the orbit, was utilized to find unseen binary companions (the so-called $O-C$ method; see, e.g. \citealt{Sterken_2005} and other papers in those proceedings). 
Such a pulse timing method works well in the case of stars pulsating with a single mode, since the intensity maxima are easy to track and any deviations from precise periodicity are fairly easy to detect. However, when the pulsating star is multiperiodic as in the case of most objects observed from space, the situation is much more complex and measurement of the time delay by careful analysis of phase modulation is more suitable.
Figure\,\ref{shibahashi:fig05} shows a time delay curve calculated with {\it Kepler} data. The time delay curve immediately provides us with qualitative information about the orbit \citep{Murphy_Shibahashi_2015, PM4_2016}.

\citet{Murphy_et_al_2018} applied this technique to all targets in the original {\it Kepler} field with effective temperatures ranging from 6600 to 10000\,K and discovered 341 new binary systems containing $\delta$ Scuti stars (main-sequence A stars pulsating in pressure modes). Importantly, many of these binaries would not have been detectable by other techniques, because A stars are often rapid rotators \citep{royeretal2007}, making spectroscopic radial velocities difficult to obtain. 
Using space-based photometry to measure the phase modulation of pulsating stars is a very efficient way to create a homogeneous sample of binary systems. 
Indeed, these asteroseismically detected binaries tripled the number of intermediate-mass binaries with full orbital solutions, and importantly, provided a homogeneous dataset for statistical analysis. 

\begin{figure}[t]
 \centering
 \includegraphics[width=0.6\textwidth,clip]{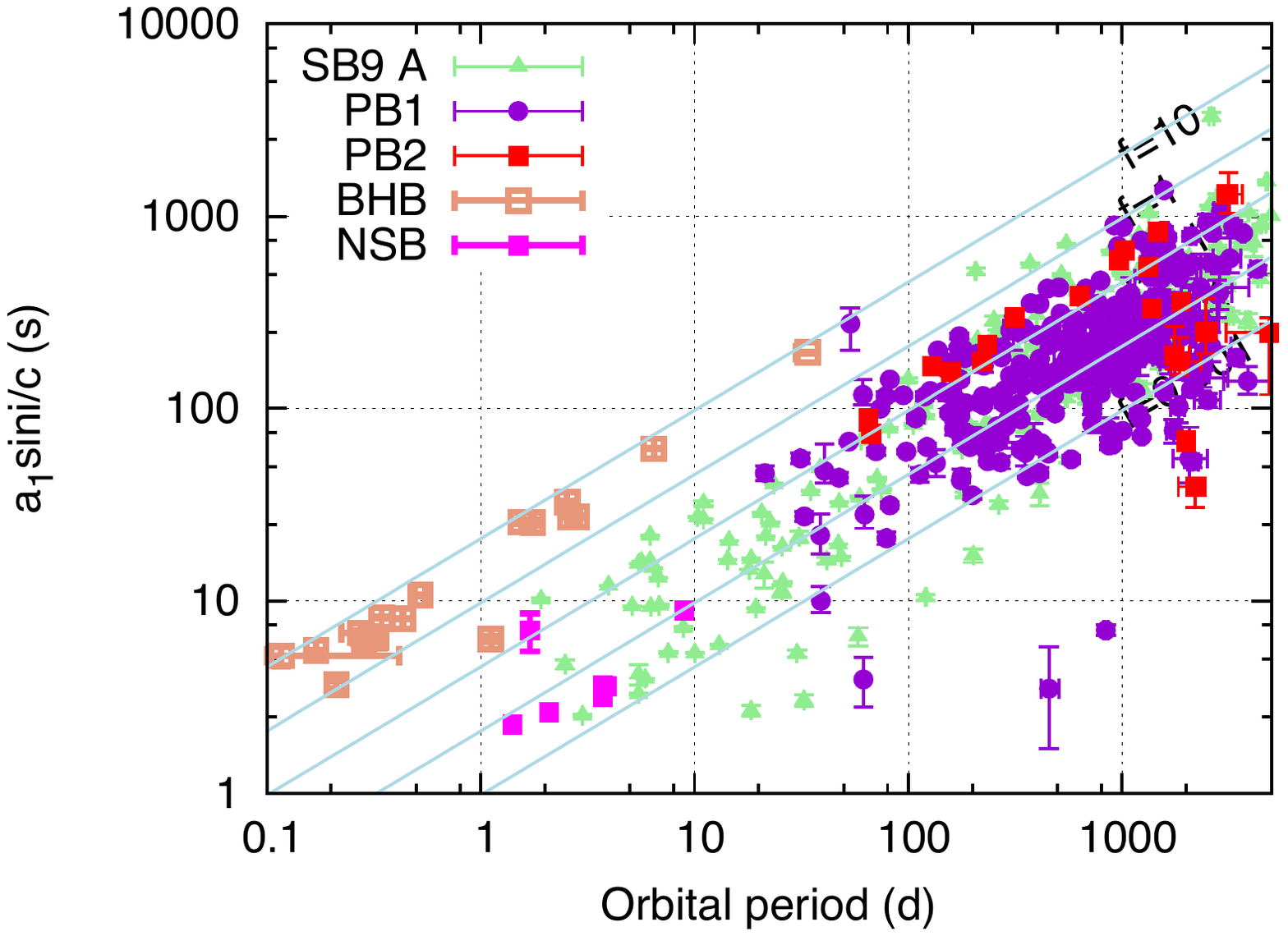}      
  \caption{Orbital period vs projected semi-major axis of newly discovered $\delta$ Sct binary systems, together with binaries with an A-type star as the primary catalogued in SB9.
  Mass functions are shown in units of ${\rm M}_\odot$.}
  \label{shibahashi:fig06}
\end{figure}

The newly detected binaries are plotted in the $(P_{\rm orb}$-$a_1\sin i/c)$-diagram shown as Fig.\,\ref{shibahashi:fig06}, together with, for comparison, the black-hole X-ray binaries listed in Table\,\ref{table:01} and 162 spectroscopic binaries listed in the ninth catalogue of spectroscopic binaries with primary stars of similar spectral type (A0--F5). Only systems with full orbital solutions with uncertainties were selected. 
Some remarks should be given here.
Binaries with orbital periods shorter than 20\,d were not found by the asteroseismic method. This is because the light curve is divided into short segments, such as 10\,d, in order to measure the phases of pulsation modes of close frequencies. 
It is then unfavourable to deal with binary stars with orbital periods shorter than the segment size dividing the observational time span, which is typically  $\sim$10\,d. 
With short-period binaries also having smaller orbits (hence smaller light travel times), the binaries with periods in the range of 20--100\,d are difficult to detect by the asteroseismic method and the sample in this period range must be considerably incomplete. It is also in this period range that binaries are most likely to eclipse, and eclipsing binaries were removed from the asteroseismic sample so as to avoid biasing the detection \citep{Murphy_et_al_2018}.

\begin{figure}[t]
 \centering
 \includegraphics[width=0.75\textwidth,clip]{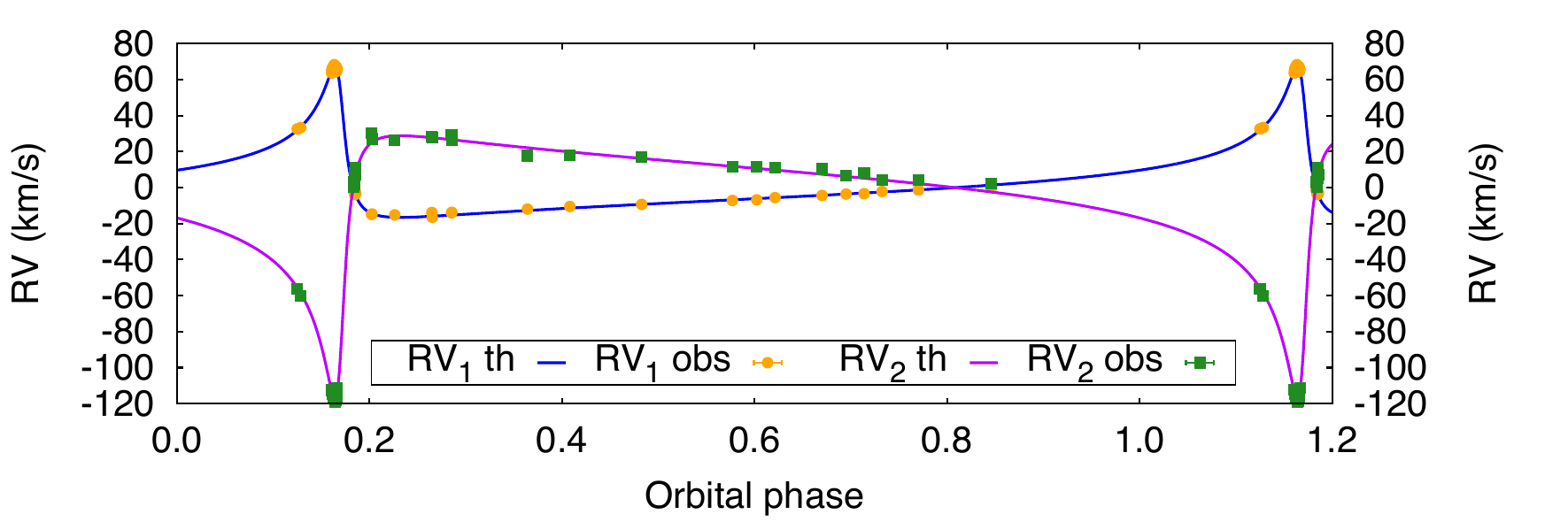}
  \caption{An example of systems with a highly eccentric orbit, about which the radial velocity observations clarified that the amplitude of phase modulation was simply overestimated.}
  \label{shibahashi:fig07}
\end{figure}

The mass range of $\delta$ Scuti type stars used by \citet{Murphy_et_al_2018} is 1.8\,$\pm$\,0.3\,M$_\odot$. Hence the systems having a mass function larger than $\sim 1\,{\rm M}_\odot$ are thought to have a binary counterpart more massive than the neutron-star mass threshold, $\sim 3\,{\rm M}_\odot$ and they may be regarded as systems containing a stellar-mass black hole.
As seen in Fig.\,\ref{shibahashi:fig06}, several systems with large mass functions have been found.
Those with a mass function larger than 
$1\,{\rm M}_\odot$ are listed in Table\,\ref{table:02}.
However, the seemingly massive secondary could be double or multiple itself, rendering each component less massive and fainter than expected. Indeed some systems with large mass functions were eventually found to be triples via follow-up with ground-based radial velocity observations made by Holger Lehman, Simon Murphy and their many collaborators (in preparation). In some other cases with a highly eccentric orbit, the radial velocity observations clarified that the amplitude of phase modulation was simply overestimated (see Fig.\,\ref{shibahashi:fig07}). 
Nevertheless, there still remain a few systems that are found to be single-line spectroscopic binary with large mass functions. They could be systems with a massive white dwarf or a neutron star. Further detailed observations are required before definitely concluding.

\renewcommand{\arraystretch}{1.2}
\begin{table}[b]
\caption{Binary properties of $\delta$\,Sct stars in the {\it Kepler} field having a mass function larger than 1\,${\rm M}_\odot$.
}
\begin{center}
\begin{tabular}{cccccc}
\toprule
KIC & $P_{\rm orb}$ (d) & $a_1\sin i/c$ (s) & $e$ & $\varpi_1$ (rad) & $f(M)$ (M$_\odot$) \\
\midrule
5219533 & $1571.0^{+15.0}_{- 13.0}$ & $1367.0^{+11.0}_{-10.0}$ & $0.5767^{+0.0081}_{-0.0081}$ & $0.478^{+0.016}_{-0.013}$ & $1.111^{+0.032}_{-0.032}$ \\
8459354 &  $53.559^{+0.020}_{-0.022}$ & $276.0^{+57.0}_{-76.0}$ & $0.9809^{+0.0062}_{-0.0180}$ & $6.085^{+0.034}_{-0.078}$ & $7.9^{+4.9}_{-6.5}$ \\
\bottomrule
\end{tabular}
\end{center}
\label{table:02}
\end{table}

\section{Self-lensing black-hole binaries}
In the case of an edge-on black-hole binary system, whose orbital axis is almost perpendicular to the line-of-sight, evidence for a lurking black hole would be available in addition to its large mass function. At superior conjunction 
when the black hole transits in front of the optical companion, contrary to the case of an ordinary eclipsing binary, a luminosity brightening is induced by the gravitational microlensing \citep{Einstein_1936, Leibovitz_Hube_1971, Maeder_1973}. In a general case, the characteristic size of this lens, called the Einstein radius $R_{\rm E}$, is given by 
\begin{equation}
	R_{\rm E} := \sqrt{
	{{4GM_{\rm BH}}\over{c^2}}{{d_{\rm S}}\over{d_{\rm L}}} \left(d_{\rm S}-d_{\rm L}\right)
	}
\label{eq:5.1}
\end{equation}
with the notation described in \citet{Gould_2000},
where $G$ is the gravitational constant, $M_{\rm BH}$ is the mass of the black hole, and
$d_{\rm S}$ and $d_{\rm L}$ denote the distance to the source and that to the lens, respectively. 
For the present case, 
$d_{\rm S} = d_{\rm L}+a(1-e^2)/(1-e\sin\varpi_{\rm opt})$, where
$a:=a_{\rm opt}+a_{\rm BH}$ is the summation of the semi-major axes of the two components and $\varpi_{\rm opt}$ is the argument of the periapsis of the optical companion. Hence the Einstein radius in units of light-seconds is
\begin{equation}
	{{R_{\rm E}}\over{c}}=4.44\times 10^{-2} \left( {{a/c}\over{100\,{\rm s}}} \right)^{1/2} 
	\left({{M_{\rm BH}}\over{{\rm M}_\odot}}\right)^{1/2} 
	\left( {{1-e^2}\over{1-e\sin\varpi_{\rm opt}}}\right)^{1/2} \,{\rm s} .
\label{eq:5.2}
\end{equation}
This is as small as a few percent of one solar radius, so the geometry for the lensing is very restricted to the edge-on case:
the inclination angle, $i$, should be in the range of  
$\uppi/2 - \left(R_{\rm opt}/a\right) \leq i \leq \uppi/2 + \left(R_{\rm opt}/a\right)$.

The tangential velocity of the black hole relative to the optical component, at the superior conjunction is
\begin{equation}
	{{v_{\rm t}}\over{c}} 
	=
	{{2\uppi (a/c)}\over{P_{\rm orb}}} {{(1-e\sin\varpi_{\rm opt})}\over{\sqrt{1-e^2}}} .
\label{eq:5.3}
\end{equation}
The duration of the transit  is then given by
\begin{eqnarray}
	t_{\rm trans} 
	&:=& 
	{{2\sqrt{R_{\rm opt}^2-a^2\cos^2 i}}\over{v_{\rm t}}} 
	\nonumber\\
	&=& 
	{{P_{\rm orb}}\over{\uppi}} 
	\left\{ \left( {{R_{\rm opt}}\over{a}} \right)^2 - \cos^2 i \right\}^{1/2}
	{ {  \sqrt{1-e^2} }\over{ 1-e\sin\varpi_{\rm opt}} } .
\label{eq:5.4}
\end{eqnarray}
Luminosity brightening repeats every superior conjunction, i.e. once per orbital period. 
Similar self-lensing systems with white dwarfs have so far been found for five cases \citep{Kruse_Agol_2014, Kawahara_et_al_2018, Masuda_et_al_2019}, so the probability to detect such a rare, but physically important event is not necessarily hopeless \citep{Masuda_Hotokezawa_2019}. 

The brightness enhancement of the microlens during the transit is $A (R_{\rm E}/R_{\rm opt})^2$, where 
we estimate $A$ to be 1.27 by integrating light emitted from points behind the Einstein radius using equation 5 of \citet{Paczynski_1986}.
Therefore, the luminosity enhancement during the black-hole transit is of the order of a few percent  in the case of $R_{\rm E}/R_{\rm opt}=1/10$. 
The transit light curves (magnification versus time) of microlensing events resemble inverted planetary transits.

\section{Search for quiet black holes by astrometry}
\begin{figure}[h!]
 \centering
 \includegraphics[angle=0,width=0.65\textwidth,clip]{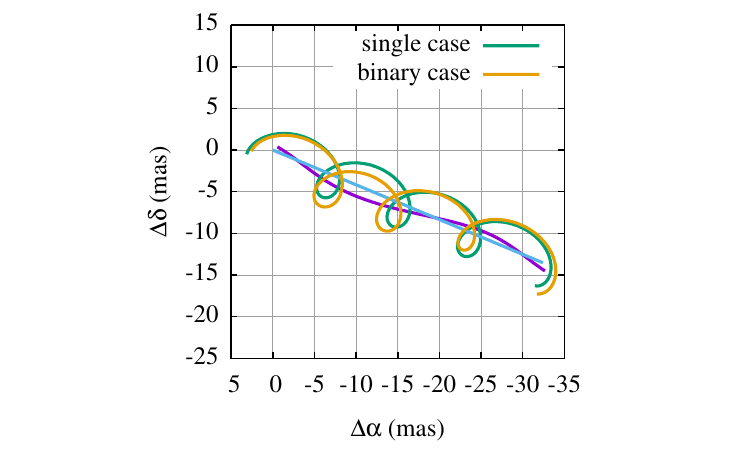}      
  \caption{Illustration of the projected motion of a single star and of a star in a binary system with an unseen companion. }
  \label{shibahashi:fig09}
\end{figure}

Another promising method of finding quiet black-hole binaries is with Gaia astrometry, which is extremely sensitive to non-linear proper motions of astrometric binaries with periods in the range 0.03 to 30 years.  
Target stars are not restricted to pulsating stars. Gaia is expected to detect ``approximately 60 per cent of the estimated 10 million binaries down to 20 mag closer than 250\,pc to the Sun''\footnote{https://sci.esa.int/web/gaia/-/31441-binary-stars}.  Like a simulation illustrated in Fig.\,\ref{shibahashi:fig09}, the orbital motion in each astrometric binary system can be seen directly, hence the orbital elements and then mass functions are determined. 
Gaia simultaneously carries out spectroscopic observations, though not for all stars. If the target has spectroscopy, then the mass of the star is independently estimated once its temperature is spectrosocpically determined. If one component is unseen, its mass is deduced from the mass function and the spectroscopic mass of the optical counterpart. The unseen single companion should be either a white dwarf, a neutron star, or a black hole, depending on its mass.

\section{Summary}
In summary,
\begin{itemize}
\item
X-ray quiet BHBs are expected to be present.
\item
Searching for X-ray quiet BHBs is challenging, but worth pursuing.
\item
Space-based asteroseismology and the measurements of light-arrival time delays opened a new window to binary statistics.
\item
Binary systems with unseen companions and high mass functions could result from the secondary being a black hole. 
\item
Self-lensing BHBs are expected to show periodic brightening.
\item
Space-based astrometry also opens another window to binary statistics. 
\end{itemize}

\section*{Acknowledgements}
\begin{acknowledgements}
The authors thank the organizing committee of the conference, led by Werner Weiss, for their excellent organization and good atmosphere of this fruitful meeting.
This work was partly supported by the JSPS Grant-in-Aid for Scientific Research (16K05288).
\end{acknowledgements}

\bibliographystyle{aa}  
\bibliography{shibahashi_5k07} 

\begin{thebibliography}{56}
\expandafter\ifx\csname natexlab\endcsname\relax\def\natexlab#1{#1}\fi

\bibitem[{{Adams} {et~al.}(2017){Adams}, {Kochanek}, {Gerke}, \&
  {Stanek}}]{Adams_et_al_2017}
{Adams}, S.~M., {Kochanek}, C.~S., {Gerke}, J.~R., \& {Stanek}, K.~Z. 2017,
  \mnras, 469, 1445

\bibitem[{{Auvergne} {et~al.}(2009){Auvergne}, {Bodin}, {Boisnard}, {Buey},
  {Chaintreuil}, {Epstein}, {Jouret}, {Lam-Trong}, {Levacher}, {Magnan},
  {Perez}, {Plasson}, {Plesseria}, {Peter}, {Steller}, {Tiph{\`e}ne}, {Baglin},
  {Agogu{\'e}}, {Appourchaux}, {Barbet}, {Beaufort}, {Bellenger}, {Berlin},
  {Bernardi}, {Blouin}, {Boumier}, {Bonneau}, {Briet}, {Butler}, {Cautain},
  {Chiavassa}, {Costes}, {Cuvilho}, {Cunha-Parro}, {de Oliveira Fialho},
  {Decaudin}, {Defise}, {Djalal}, {Docclo}, {Drummond}, {Dupuis}, {Exil},
  {Faur{\'e}}, {Gaboriaud}, {Gamet}, {Gavalda}, {Grolleau}, {Gueguen},
  {Guivarc'h}, {Guterman}, {Hasiba}, {Huntzinger}, {Hustaix}, {Imbert},
  {Jeanville}, {Johlander}, {Jorda}, {Journoud}, {Karioty}, {Kerjean},
  {Lafond}, {Lapeyrere}, {Landiech}, {Larqu{\'e}}, {Laudet}, {Le Merrer},
  {Leporati}, {Leruyet}, {Levieuge}, {Llebaria}, {Martin}, {Mazy}, {Mesnager},
  {Michel}, {Moalic}, {Monjoin}, {Naudet}, {Neukirchner}, {Nguyen-Kim},
  {Ollivier}, {Orcesi}, {Ottacher}, {Oulali}, {Parisot}, {Perruchot},
  {Piacentino}, {Pinheiro da Silva}, {Platzer}, {Pontet}, {Pradines},
  {Quentin}, {Rohbeck}, {Rolland}, {Rollenhagen}, {Romagnan}, {Russ}, {Samadi},
  {Schmidt}, {Schwartz}, {Sebbag}, {Smit}, {Sunter}, {Tello}, {Toulouse},
  {Ulmer}, {Vandermarcq}, {Vergnault}, {Wallner}, {Waultier}, \&
  {Zanatta}}]{CoRoT_2009}
{Auvergne}, M., {Bodin}, P., {Boisnard}, L., {et~al.} 2009, \aap, 506, 411

\bibitem[{{Barthelmy} {et~al.}(2005){Barthelmy}, {Barbier}, {Cummings},
  {Fenimore}, {Gehrels}, {Hullinger}, {Krimm}, {Markwardt}, {Palmer},
  {Parsons}, {Sato}, {Suzuki}, {Takahashi}, {Tashiro}, \& {Tueller}}]{BAT_2005}
{Barthelmy}, S.~D., {Barbier}, L.~M., {Cummings}, J.~R., {et~al.} 2005, \ssr,
  120, 143

\bibitem[{{Batten}(1967)}]{SB6_1967}
{Batten}, A.~H. 1967, Publications of the Dominion Astrophysical Observatory
  Victoria, 13, 119

\bibitem[{{Bolton}(1972)}]{Bolton_1972}
{Bolton}, C.~T. 1972, Nature Physical Science, 240, 124

\bibitem[{{Breivik} {et~al.}(2017){Breivik}, {Chatterjee}, \&
  {Larson}}]{Breivik_et_al_2017}
{Breivik}, K., {Chatterjee}, S., \& {Larson}, S.~L. 2017, \apjl, 850, L13

\bibitem[{{Brown} \& {Bethe}(1994)}]{Brown_Bethe_1994}
{Brown}, G.~E. \& {Bethe}, H.~A. 1994, \apj, 423, 659

\bibitem[{{Casares} {et~al.}(1992){Casares}, {Charles}, \&
  {Naylor}}]{Casares_et_al_1992}
{Casares}, J., {Charles}, P.~A., \& {Naylor}, T. 1992, \nat, 355, 614

\bibitem[{{Casares} \& {Jonker}(2014)}]{Casares_Jonker_2014}
{Casares}, J. \& {Jonker}, P.~G. 2014, \ssr, 183, 223

\bibitem[{{Casares} {et~al.}(2019){Casares}, {Mu{\~n}oz-Darias}, {Mata
  S{\'a}nchez}, {Charles}, {Torres}, {Armas Padilla}, {Fender}, \&
  {Garc{\'\i}a-Rojas}}]{Casares_et_al_2019}
{Casares}, J., {Mu{\~n}oz-Darias}, T., {Mata S{\'a}nchez}, D., {et~al.} 2019,
  \mnras, 488, 1356

\bibitem[{{Corral-Santana} {et~al.}(2016){Corral-Santana}, {Casares},
  {Mu{\~n}oz-Darias}, {Bauer}, {Mart{\'\i}nez-Pais}, \&
  {Russell}}]{Corral-Santana_et_al_2016}
{Corral-Santana}, J.~M., {Casares}, J., {Mu{\~n}oz-Darias}, T., {et~al.} 2016,
  \aap, 587, A61

\bibitem[{{Cowley}(1992)}]{Cowley_1992}
{Cowley}, A.~P. 1992, \araa, 30, 287

\bibitem[{{Einstein}(1936)}]{Einstein_1936}
{Einstein}, A. 1936, Science, 84, 506

\bibitem[{{Gehrels} {et~al.}(2004){Gehrels}, {Chincarini}, {Giommi}, {Mason},
  {Nousek}, {Wells}, {White}, {Barthelmy}, {Burrows}, {Cominsky}, {Hurley},
  {Marshall}, {M{\'e}sz{\'a}ros}, {Roming}, {Angelini}, {Barbier}, {Belloni},
  {Campana}, {Caraveo}, {Chester}, {Citterio}, {Cline}, {Cropper}, {Cummings},
  {Dean}, {Feigelson}, {Fenimore}, {Frail}, {Fruchter}, {Garmire}, {Gendreau},
  {Ghisellini}, {Greiner}, {Hill}, {Hunsberger}, {Krimm}, {Kulkarni}, {Kumar},
  {Lebrun}, {Lloyd-Ronning}, {Markwardt}, {Mattson}, {Mushotzky}, {Norris},
  {Osborne}, {Paczynski}, {Palmer}, {Park}, {Parsons}, {Paul}, {Rees},
  {Reynolds}, {Rhoads}, {Sasseen}, {Schaefer}, {Short}, {Smale}, {Smith},
  {Stella}, {Tagliaferri}, {Takahashi}, {Tashiro}, {Townsley}, {Tueller},
  {Turner}, {Vietri}, {Voges}, {Ward}, {Willingale}, {Zerbi}, \&
  {Zhang}}]{Swift_2004}
{Gehrels}, N., {Chincarini}, G., {Giommi}, P., {et~al.} 2004, \apj, 611, 1005

\bibitem[{{Gould}(2000)}]{Gould_2000}
{Gould}, A. 2000, \apj, 542, 785

\bibitem[{{Guseinov} \& {Zel'dovich}(1966)}]{Guseinov_Zeldovich_1966}
{Guseinov}, O.~K. \& {Zel'dovich}, Y.~B. 1966, Soviet Astronomy, 10, 251
  (translated from Astronomicheskii Zhurnal, 43, 313)

\bibitem[{{Heger} {et~al.}(2003){Heger}, {Fryer}, {Woosley}, {Langer}, \&
  {Hartmann}}]{Heger_et_al_2003}
{Heger}, A., {Fryer}, C.~L., {Woosley}, S.~E., {Langer}, N., \& {Hartmann},
  D.~H. 2003, \apj, 591, 288

\bibitem[{{Kawahara} {et~al.}(2018){Kawahara}, {Masuda}, {MacLeod}, {Latham},
  {Bieryla}, \& {Benomar}}]{Kawahara_et_al_2018}
{Kawahara}, H., {Masuda}, K., {MacLeod}, M., {et~al.} 2018, \aj, 155, 144

\bibitem[{{Kirk} {et~al.}(2016){Kirk}, {Conroy}, {Pr{\v{s}}a}, {Abdul-Masih},
  {Kochoska}, {Matijevi{\v{c}}}, {Hambleton}, {Barclay}, {Bloemen}, {Boyajian},
  {Doyle}, {Fulton}, {Hoekstra}, {Jek}, {Kane}, {Kostov}, {Latham}, {Mazeh},
  {Orosz}, {Pepper}, {Quarles}, {Ragozzine}, {Shporer}, {Southworth},
  {Stassun}, {Thompson}, {Welsh}, {Agol}, {Derekas}, {Devor}, {Fischer},
  {Green}, {Gropp}, {Jacobs}, {Johnston}, {LaCourse}, {Saetre}, {Schwengeler},
  {Toczyski}, {Werner}, {Garrett}, {Gore}, {Martinez}, {Spitzer}, {Stevick},
  {Thomadis}, {Vrijmoet}, {Yenawine}, {Batalha}, \&
  {Borucki}}]{Kirk_et_al_2016}
{Kirk}, B., {Conroy}, K., {Pr{\v{s}}a}, A., {et~al.} 2016, \aj, 151, 68

\bibitem[{{Koch} {et~al.}(2010){Koch}, {Borucki}, {Basri}, {Batalha}, {Brown},
  {Caldwell}, {Christensen-Dalsgaard}, {Cochran}, {DeVore}, {Dunham},
  {Gautier}, {Geary}, {Gilliland}, {Gould}, {Jenkins}, {Kondo}, {Latham},
  {Lissauer}, {Marcy}, {Monet}, {Sasselov}, {Boss}, {Brownlee}, {Caldwell},
  {Dupree}, {Howell}, {Kjeldsen}, {Meibom}, {Morrison}, {Owen}, {Reitsema},
  {Tarter}, {Bryson}, {Dotson}, {Gazis}, {Haas}, {Kolodziejczak}, {Rowe}, {Van
  Cleve}, {Allen}, {Chand rasekaran}, {Clarke}, {Li}, {Quintana}, {Tenenbaum},
  {Twicken}, \& {Wu}}]{Kepler_2010}
{Koch}, D.~G., {Borucki}, W.~J., {Basri}, G., {et~al.} 2010, \apjl, 713, L79

\bibitem[{{Kruse} \& {Agol}(2014)}]{Kruse_Agol_2014}
{Kruse}, E. \& {Agol}, E. 2014, Science, 344, 275

\bibitem[{{Lamberts} {et~al.}(2018){Lamberts}, {Garrison-Kimmel}, {Hopkins},
  {Quataert}, {Bullock}, {Faucher-Gigu{\`e}re}, {Wetzel}, {Kere{\v{s}}},
  {Drango}, \& {Sand erson}}]{Lamberts_et_al_2018}
{Lamberts}, A., {Garrison-Kimmel}, S., {Hopkins}, P.~F., {et~al.} 2018, \mnras,
  480, 2704

\bibitem[{{Leibovitz} \& {Hube}(1971)}]{Leibovitz_Hube_1971}
{Leibovitz}, C. \& {Hube}, D.~P. 1971, \aap, 15, 251

\bibitem[{{Maeder}(1973)}]{Maeder_1973}
{Maeder}, A. 1973, \aap, 26, 215

\bibitem[{{Majewski} {et~al.}(2017){Majewski}, {Schiavon}, {Frinchaboy},
  {Allende Prieto}, {Barkhouser}, {Bizyaev}, {Blank}, {Brunner}, {Burton},
  {Carrera}, {Chojnowski}, {Cunha}, {Epstein}, {Fitzgerald}, {Garc{\'\i}a
  P{\'e}rez}, {Hearty}, {Henderson}, {Holtzman}, {Johnson}, {Lam}, {Lawler},
  {Maseman}, {M{\'e}sz{\'a}ros}, {Nelson}, {Nguyen}, {Nidever}, {Pinsonneault},
  {Shetrone}, {Smee}, {Smith}, {Stolberg}, {Skrutskie}, {Walker}, {Wilson},
  {Zasowski}, {Anders}, {Basu}, {Beland}, {Blanton}, {Bovy}, {Brownstein},
  {Carlberg}, {Chaplin}, {Chiappini}, {Eisenstein}, {Elsworth}, {Feuillet},
  {Fleming}, {Galbraith-Frew}, {Garc{\'\i}a}, {Garc{\'\i}a-Hern{\'a}ndez},
  {Gillespie}, {Girardi}, {Gunn}, {Hasselquist}, {Hayden}, {Hekker}, {Ivans},
  {Kinemuchi}, {Klaene}, {Mahadevan}, {Mathur}, {Mosser}, {Muna}, {Munn},
  {Nichol}, {O'Connell}, {Parejko}, {Robin}, {Rocha-Pinto}, {Schultheis},
  {Serenelli}, {Shane}, {Silva Aguirre}, {Sobeck}, {Thompson}, {Troup},
  {Weinberg}, \& {Zamora}}]{APOGEE_2017}
{Majewski}, S.~R., {Schiavon}, R.~P., {Frinchaboy}, P.~M., {et~al.} 2017, \aj,
  154, 94

\bibitem[{{Mashian} \& {Loeb}(2017)}]{Mashian_Abraham_2017}
{Mashian}, N. \& {Loeb}, A. 2017, \mnras, 470, 2611

\bibitem[{{Masuda} \& {Hotokezaka}(2019)}]{Masuda_Hotokezawa_2019}
{Masuda}, K. \& {Hotokezaka}, K. 2019, \apj, 883, 169

\bibitem[{{Masuda} {et~al.}(2019){Masuda}, {Kawahara}, {Latham}, {Bieryla},
  {Kunitomo}, {MacLeod}, \& {Aoki}}]{Masuda_et_al_2019}
{Masuda}, K., {Kawahara}, H., {Latham}, D.~W., {et~al.} 2019, \apjl, 881, L3

\bibitem[{{Mineshige} \& {Wheeler}(1989)}]{Mineshige_Wheeler_1989}
{Mineshige}, S. \& {Wheeler}, J.~C. 1989, \apj, 343, 241

\bibitem[{{Murphy}(2018)}]{Murphy_2018}
{Murphy}, S.~J. 2018, arXiv e-prints, arXiv:1811.12659

\bibitem[{{Murphy} {et~al.}(2014){Murphy}, {Bedding}, {Shibahashi}, {Kurtz}, \&
  {Kjeldsen}}]{PM_2014}
{Murphy}, S.~J., {Bedding}, T.~R., {Shibahashi}, H., {Kurtz}, D.~W., \&
  {Kjeldsen}, H. 2014, \mnras, 441, 2515

\bibitem[{{Murphy} {et~al.}(2018){Murphy}, {Moe}, {Kurtz}, {Bedding},
  {Shibahashi}, \& {Boffin}}]{Murphy_et_al_2018}
{Murphy}, S.~J., {Moe}, M., {Kurtz}, D.~W., {et~al.} 2018, \mnras, 474, 4322

\bibitem[{{Murphy} \& {Shibahashi}(2015)}]{Murphy_Shibahashi_2015}
{Murphy}, S.~J. \& {Shibahashi}, H. 2015, \mnras, 450, 4475

\bibitem[{{Murphy} {et~al.}(2016){Murphy}, {Shibahashi}, \&
  {Bedding}}]{PM4_2016}
{Murphy}, S.~J., {Shibahashi}, H., \& {Bedding}, T.~R. 2016, \mnras, 461, 4215

\bibitem[{{Paczy{\'n}ski}(1971)}]{Paczynski_1971}
{Paczy{\'n}ski}, B. 1971, \araa, 9, 183

\bibitem[{{Paczynski}(1986)}]{Paczynski_1986}
{Paczynski}, B. 1986, \apj, 304, 1

\bibitem[{{Pourbaix} {et~al.}(2004){Pourbaix}, {Tokovinin}, {Batten}, {Fekel},
  {Hartkopf}, {Levato}, {Morrell}, {Torres}, \& {Udry}}]{SB9_2004}
{Pourbaix}, D., {Tokovinin}, A.~A., {Batten}, A.~H., {et~al.} 2004, \aap, 424,
  727

\bibitem[{{Remillard} \& {McClintock}(2006)}]{Remillard_McClintock_2006}
{Remillard}, R.~A. \& {McClintock}, J.~E. 2006, \araa, 44, 49

\bibitem[{{Ricker} {et~al.}(2015){Ricker}, {Winn}, {Vanderspek}, {Latham},
  {Bakos}, {Bean}, {Berta-Thompson}, {Brown}, {Buchhave}, {Butler}, {Butler},
  {Chaplin}, {Charbonneau}, {Christensen-Dalsgaard}, {Clampin}, {Deming},
  {Doty}, {De Lee}, {Dressing}, {Dunham}, {Endl}, {Fressin}, {Ge}, {Henning},
  {Holman}, {Howard}, {Ida}, {Jenkins}, {Jernigan}, {Johnson}, {Kaltenegger},
  {Kawai}, {Kjeldsen}, {Laughlin}, {Levine}, {Lin}, {Lissauer}, {MacQueen},
  {Marcy}, {McCullough}, {Morton}, {Narita}, {Paegert}, {Palle}, {Pepe},
  {Pepper}, {Quirrenbach}, {Rinehart}, {Sasselov}, {Sato}, {Seager},
  {Sozzetti}, {Stassun}, {Sullivan}, {Szentgyorgyi}, {Torres}, {Udry}, \&
  {Villasenor}}]{TESS_2015}
{Ricker}, G.~R., {Winn}, J.~N., {Vanderspek}, R., {et~al.} 2015, Journal of
  Astronomical Telescopes, Instruments, and Systems, 1, 014003

\bibitem[{{Royer} {et~al.}(2007){Royer}, {Zorec}, \&
  {G{\'o}mez}}]{royeretal2007}
{Royer}, F., {Zorec}, J., \& {G{\'o}mez}, A.~E. 2007, \aap, 463, 671

\bibitem[{{Shibahashi} \& {Kurtz}(2012)}]{FM_2012}
{Shibahashi}, H. \& {Kurtz}, D.~W. 2012, \mnras, 422, 738

\bibitem[{{Shibahashi} {et~al.}(2015){Shibahashi}, {Kurtz}, \&
  {Murphy}}]{FM2_2015}
{Shibahashi}, H., {Kurtz}, D.~W., \& {Murphy}, S.~J. 2015, \mnras, 450, 3999

\bibitem[{{Shibahashi} \& {Murphy}(2019)}]{Liege_2019}
{Shibahashi}, H. \& {Murphy}, S.~J. 2019, arXiv e-prints, arXiv:1909.07595

\bibitem[{{Sterken}(2005)}]{Sterken_2005}
{Sterken}, C. 2005, The Light-Time Effect in Astrophysics: Causes and cures of
  the O-C diagram, ASP Conference Series, 335

\bibitem[{{Stickland}(1997)}]{Stickland_1997}
{Stickland}, D.~J. 1997, The Observatory, 117, 37

\bibitem[{{Tanaka} \& {Shibazaki}(1996)}]{Tanaka_Shibazaki_1996}
{Tanaka}, Y. \& {Shibazaki}, N. 1996, \araa, 34, 607

\bibitem[{{Thompson} {et~al.}(2019){Thompson}, {Kochanek}, {Stanek}, {Badenes},
  {Post}, {Jayasinghe}, {Latham}, {Bieryla}, {Esquerdo}, {Berlind}, {Calkins},
  {Tayar}, {Lindegren}, {Johnson}, {Holoien}, {Auchettl}, \&
  {Covey}}]{Thompson_2019}
{Thompson}, T.~A., {Kochanek}, C.~S., {Stanek}, K.~Z., {et~al.} 2019, Science,
  366, 637

\bibitem[{{Torres} {et~al.}(2019){Torres}, {Casares}, {Jim{\'e}nez-Ibarra},
  {Mu{\~n}oz-Darias}, {Armas Padilla}, {Jonker}, \&
  {Heida}}]{Torres_et_al_2019}
{Torres}, M.~A.~P., {Casares}, J., {Jim{\'e}nez-Ibarra}, F., {et~al.} 2019,
  \apjl, 882, L21

\bibitem[{{Trimble} \& {Thorne}(2018)}]{Trimble_Thorne_2018}
{Trimble}, V. \& {Thorne}, K.~S. 2018, arXiv e-prints, arXiv:1811.04310

\bibitem[{{Trimble} \& {Thorne}(1969)}]{Trimble_Thorne_1969}
{Trimble}, V.~L. \& {Thorne}, K.~S. 1969, \apj, 156, 1013

\bibitem[{{Walker} {et~al.}(2003){Walker}, {Matthews}, {Kuschnig}, {Johnson},
  {Rucinski}, {Pazder}, {Burley}, {Walker}, {Skaret}, {Zee}, {Grocott},
  {Carroll}, {Sinclair}, {Sturgeon}, \& {Harron}}]{MOST_2003}
{Walker}, G., {Matthews}, J., {Kuschnig}, R., {et~al.} 2003, \pasp, 115, 1023

\bibitem[{{Webster} \& {Murdin}(1972)}]{Webster_Murdin_1972}
{Webster}, B.~L. \& {Murdin}, P. 1972, \nat, 235, 37

\bibitem[{{Weiss} {et~al.}(2014){Weiss}, {Rucinski}, {Moffat},
  {Schwarzenberg-Czerny}, {Koudelka}, {Grant}, {Zee}, {Kuschnig}, {Mochnacki},
  {Matthews}, {Orleanski}, {Pamyatnykh}, {Pigulski}, {Alves}, {Guedel},
  {Handler}, {Wade}, \& {Zwintz}}]{BRITE_2014}
{Weiss}, W.~W., {Rucinski}, S.~M., {Moffat}, A.~F.~J., {et~al.} 2014, \pasp,
  126, 573

\bibitem[{{Wheeler}(1968)}]{Wheeler_1968}
{Wheeler}, J.~A. 1968, American Scientist, 56, 1

\bibitem[{{Yalinewich} {et~al.}(2018){Yalinewich}, {Beniamini}, {Hotokezaka},
  \& {Zhu}}]{Yalinewich_et_al_2018}
{Yalinewich}, A., {Beniamini}, P., {Hotokezaka}, K., \& {Zhu}, W. 2018, \mnras,
  481, 930

\bibitem[{{Yamaguchi} {et~al.}(2018){Yamaguchi}, {Kawanaka}, {Bulik}, \&
  {Piran}}]{Yamaguchi_et_al_2018}
{Yamaguchi}, M.~S., {Kawanaka}, N., {Bulik}, T., \& {Piran}, T. 2018, \apj,
  861, 21

\end{thebibliography}

\end{document}